# Critical Current Density of Advanced Internal-Mg-Diffusion-Processed $MgB_2$ Wires

G Z Li[1], M D Sumption[1], M A Susner[1], Y Yang[1], K M Reddy[1], M A Rindfleisch[2], M J Tomsic[2], C J Thong[2], and E W Collings[1]

[1] Center for Superconducting and Magnetic Materials, Department of Materials Science and Engineering, the Ohio State University, Columbus, OH 43210, U.S.A.

[2] Hyper Tech Research Incorporated, Columbus, OH 43212, U.S.A.

E-mail: li.1423@osu.edu

**Abstract**

Recent advances in $MgB_2$ conductors are leading to a new level of performance. Based on the use of proper powders, proper chemistry, and an architecture which incorporates internal Mg diffusion (IMD), a dense $MgB_2$ structure with not only a high critical current density $J_c$, but also a high engineering critical current density, $J_e$, can be obtained. In this paper, a series of these advanced (or second-generation, "2G") conductors has been prepared. Scanning electron microscopy and associated energy dispersive X-ray spectroscopy were applied to characterize the microstructures and compositions of the wires, and a dense $MgB_2$ layer structure was observed. The best layer $J_c$ for our sample is $1.07 \times 10^5$ A/cm$^2$ at 10 T, 4.2 K, and our best $J_e$ is seen to be $1.67 \times 10^4$ A/cm$^2$ at 10 T, 4.2 K. Optimization of the transport properties of these advanced wires is discussed in terms of B-powder choice, area fraction, and the $MgB_2$ layer growth mechanism.

# 1. Introduction

The discovery of MgB$_2$, with its critical temperature $T_c$, of about 39 K, and hence the ability to operate at temperatures beyond the range of the low temperature superconductors NbTi and Nb$_3$Sn encouraged many researchers to develop MgB$_2$ as a practical superconducting wire [1]. Such wires have been made using two popular powder-in-tube (PIT) approaches – *ex-situ* [2, 3] and *in-situ* [4, 5]. Over the years, numerous efforts have been made to improve the pinning properties [6] and solve the connectivity and porosity issues of MgB$_2$ wires [7] for the purpose of obtaining high critical current densities. An interesting variant of the *in-situ* PIT route is the "internal Mg diffusion" (IMD) process or the "Mg-reactive liquid infiltration" (Mg RLI) process. Initiated by Giunchi *et al.* [8, 9], IMD wire formation starts off with a Mg rod embedded axially in a B-filled tube and continues with wire drawing and final heat treatment, HT [10, 11]. Compared to the conventional *in-situ* PIT process which yields a system of randomly connected MgB$_2$ fibers associated with copious porosity, the IMD process can produce a dense MgB$_2$ layer structure with excellent longitudinal and transverse connectivities [12]. The transport properties of IMD wires are parameterized in terms of (i) the "layer, $J_c$" (herein designated $J_c$) which is the critical transport current, $I_c$, divided by the cross-sectional area of the MgB$_2$ layer, and (ii) the "engineering critical current density, $J_e$," which is $I_c$ divided by the cross-sectional area of the whole strand.

Kumakura *et al.* fabricated a series of SiC doped mono- and multi- filament MgB$_2$ wires using the IMD process and applied various HT conditions [13]. Their wires attained a high $J_c$ of $1.1 \times 10^5$ A/cm$^2$ 4.2 K and 10 T. However, the cross sectional area of the reacted layer was so low that its $I_c$, only about 10 A, translated into a very small value of $J_e$.

Subsequently Ye *et al.* by doping their IMD-precursor wires with both SiC and a liquid aromatic hydrocarbon obtained a 4.2 K, 10 T $J_e$ of $3.7\times10^3$ A/cm$^2$ [14]. Our variant of the IMD approach, which included fine grain B and C-doping produced very high layer $J_c$s and $n$ values [12]. In continuing along these lines we have optimized the strand architecture to obtain higher $J_e$ values as well. Indeed, combining the IMD process with optimal powders, doping, and strand architectures leads to such substantial improvements in MgB$_2$ performance that we are proposing to name the new conductor "Second Generation MgB$_2$". In what follows we describe a series of monofilamentary IMD wires and the various HT conditions that were applied to optimize their layer $J_c$s and engineering $J_e$s. Based on the structural and transport results, the MgB$_2$ layer growth mechanism in IMD wires as well as several parameters that affect the final $J_e$ are discussed.

## 2. Experimental

A series of precursor monofilamentary strands with a Nb barrier and a monel outer sheath was manufactured by Hyper Tech Research, Inc. (HTR). The initial wire billet is a Mg rod positioned along the axis of a B-filled double tube of Nb and monel. All the samples used fine (10-100 nm) mostly amorphous B powder from Specialty Materials Inc. produced by the plasma assisted reaction of BCl$_3$ with H$_2$ and *in situ* C doped (2 mol%) by adding CH$_4$ to the process gases [15-18]. The reactivity of such small particles with their large surface/volume ratios favors the formation of thick MgB$_2$ reaction layers. The influence of B type (size) on the transport properties of PIT MgB$_2$ is described in [19]. The billet was drawn to 0.83 mm OD and 0.55 mm OD wires. During HT in a tube furnace the wires were ramped to soak temperature in about 80 min, kept at 675 °C for up to 8 hours and furnace cooled to room temperature. The HT temperature of 675 ºC was chosen based on extensive experience with the HT of CTFF-processed PIT wires, as a result of which we found little variation in $J_c$ with HT at 650 ºC to 700 ºC for suitable times. In fact, 675 ºC is often regarded as the optimal temperature for CTFF wires. A recent paper on such wires compared the results of HT at 600 ºC and 700 ºC [20]. We found that at 600 ºC the Mg + B → $MgB_2$ reaction required at least 71 h, while at 700 ºC the reaction was already complete after 0.5 h. Prolonged HT at 700 ºC exposes the forming $MgB_2$ to contamination by Cu leaking from the monel sheath through the overlap-seamed Nb chemical barrier. For these reasons we chose 675 ºC as the HT temperature for our present IMD wires. This yielded a satisfactory result in that strand B3, for example, was fully transformed (no excess B after 4 h at 675 ºC) into a hollow cylinder of $MgB_2$ surrounding the site of the starting Mg rod. The strand specifications and HT conditions are listed in Table I.

Transport measurements of $J_c$ were carried out in transverse magnetic fields, $B$, of up to 12 T in pool boiling liquid He at 4.2 K on samples 50 mm long. The voltage-tap separation (gauge length) distance of 5 mm provided a gauge-length/sample-length ratio of 1:10, and voltage-tap to current-contact separation distances were well above the current-transfer lengths of 5-10 mm, thereby ensuring complete current sharing into the $MgB_2$ layers. The critical current was determined at an electric field criterion of 1 $\mu$V/cm.

Microstructures and compositions of the $MgB_2$ wires were characterized by scanning electron microscopy (SEM) and energy dispersive X-ray spectroscopy (EDS). SEM observations were carried out using an FEI Sirion FEG SEM and a Quanta 200 SEM equipped with the EDAX EDS system.

## 3. Results

*3.1. Microstructure of the present wires*

Figure 1 shows the back scattered (BSE) SEM images of the 0.83 mm diameter A-series samples. Sharp interfaces are formed between adjacent layers. The Mg is not fully reacted even after 8 hours HT. Some of it remains in the central core region previously occupied by the starting Mg rod. The previous B layer has been transformed into two separate layers. EDS analysis has been employed to characterize the Mg: B atomic ratios of these two layers. The Mg: B ratios of the inner and outer layers are about 1:2 and 1:7 respectively, indicating that the inner layer is fully transformed $MgB_2$ and the outer layer is a partially reacted "B-rich region" where a relatively small amount of B has reacted with Mg. This result has been verified by the high resolution fracture-SEM images of figure 2. Figure 2 exhibits the microstructural difference between the two layers. The $MgB_2$ layer appears to be uniformly dense. However the B rich layer has the appearance of ball-shaped packs of powder, much like the microstructure of the original B powder before HT. The $MgB_2$ layer thicknesses and areas are listed in Table II, in which we note that thicknesses of 30-60 μm (A series) and 40-90 μm (B series) have been achieved. Layer thickness is influenced by many factors including the B powder particle size and packing density. In SiC doped IMD samples (B: 99.9%, 300 mesh, 40-50 μm), Kumakura noted that the diffusion distance of Mg into the doped B layer was limited to about 20-30 μm [21]. Our B powder from SMI is "fluffy", fine (~40 nm) and of low density. As a result, the Mg is able to percolate into it further than it does in denser granular powders (e.g. 300 mesh [21]). For this reason, in the present strands thicknesses of 30-60 μm (A series) and 40-90 μm (B series) have been achieved. Strand B3 is fully reacted; because of the eccentricity of the Mg rod, the reaction layer thickness was caused to vary from almost zero on the "pinch" side to about 90 μm on the opposite side. Likewise in the sister B-series strands HT for 0.5 h (B1) and 1.0 h (B2) the reaction layer thickness varies from 0-40 μm (B1) and 0-49 μm (B2). Figure 3 shows the SEM images of the 0.55 mm diameter B-series samples. In general, in all the strands, the Mg rod was slightly off center; and we note that the $MgB_2$ thickness was slight less on the "pinch" side than on the opposite side. We suggest this could be caused by small azimuthal variations in B powder density.

*3.2. Transport layer $J_c$*

The layer $J_c$s of the present $MgB_2$ wires at 4.2 K in fields around 9-12 T are presented in figure 4. The cross sectional areas of our $MgB_2$ layers as measured on the respective SEM images are listed in Table II. Figure 4 shows that:

(i) The layer $J_c$s of the large diameter (0.83 mm OD) A-series wires increase strongly with HT time from 30 min to 8 hours. Sample A3 (HT for 8 hours) has the largest layer $J_c$ of the series, for example $1.04 \times 10^5$ A/cm$^2$ at 10 T. Sample A2 exhibits a "lower field" drop-off in $J_c$ as a result of sample heating; the lack of stability of this particular sample is presumably the result of relatively high currents in combination with thermal isolation of the $MgB_2$ layer by the surrounding unreacted B powder, figure 1 (b).

(ii) The layer $J_c$s of the smaller diameter (0.55 mm OD) B-series wires actually decrease slightly (13%) as the HT time increases from 30 min to 4 hours. Thus sample B1 (HT for 30 min) exhibits the greatest 10 T layer $J_c$ of the set, viz. $1.07 \times 10^5$ A/cm$^2$.

(iii) In general, the lack of stability of all the strands causes them to quench when the magnetic field drops below 9 T and 4.2 K. As a result, no meaningful $J_c$ transitions are obtainable at fields less than 9 T.

Finally, we note that the layer $J_c$s of sample A3 and all the B-series strands are practically identical to those of an IMD sample previously measured in our laboratory[12]. The results of this previous study also offer a comparison of the magnetic and transport IMD layer $J_c$s, as well as a detailed comparison of the transport properties of a pair of PIT and IMD strands.

*3.3. Formation of the MgB$_2$ layer*

The formation and growth of the MgB$_2$ layer can be described as a three-step process: (1) Melting of the Mg rod followed by rapid infiltration of the resulting liquid into the surrounding B powder. In a relatively short time the Mg can penetrate several micrometers into the surrounding B powder. (2) Reaction with the B particles. As the HT proceeds, more and more B particles transform into MgB$_2$ grains which continue to grow and coalesce. Since the molar volume of MgB$_2$ (17.46 cm$^3$/mol) is 4 times larger than that of B (4.59 cm$^3$/mol) [22], the interstices between the initial B powder particles become filled with MgB$_2$ as the reaction proceeds. As a result, a MgB$_2$ layer with almost 100% density is formed close to the surface of the prior Mg rod (now liquid Mg). However, this layer is too dense to permit further liquid Mg infiltration. (3) Reaction with the remaining B, during which Mg atoms have to diffuse through this dense MgB$_2$ layer. Step (3) is much slower than steps (1) and (2). As a result, the MgB$_2$ layer growth slows down after the fast growth of the initial layer, as confirmed by SEM images taken after HT for successively increasing periods of time.

*3.4. Layer $J_c$, wire diameter, and HT time*

The $MgB_2$ layer formation mechanisms described above are responsible for the changes in layer $J_c$ with HT time in our two classes of wire, already noted in Section 3.2 and figure 4.

For the larger diameter (0.83 mm OD) A-series wires the growth of the $MgB_2$ layer with time at 675 °C is depicted in figure 1. After 30 minutes the $MgB_2$ layer that has begun to form contains many unreacted B particles. As a result the measured layer $J_c$ is relatively small – only $0.35 \times 10^5$ A/cm$^2$, at 4.2 K, 10 T. After 1 hour of HT about 32% of the B has been consumed and after 8 hours 39% of the B is reacted. The actual layer $J_c$s of these two wires are the same at high fields; the high values obtained (~ $10^5$ A/cm$^2$ at 10 T, 4.2 K) indicating that dense uniform layers had been formed in both cases. The instability of strand A2 at fields below 11 T was referred to in Section 3.2.

For the smaller diameter (0.55 mm OD) B-series wires a dense $MgB_2$ layer has already begun to form after just 30 min of HT, figure 3(a). The reaction time is shorter mainly because the total distance that the Mg must be transported through the B layer is shorter, although there are some indications that the $MgB_2$ layer may be growing faster as well. As the HT time increases to 1 hour and then to 4 hours the reaction layer gets thicker but its layer-$J_c$ undergoes a small monotonic decrease in response to grain growth.

*3.5. Engineering Critical Current Density, $J_e$*

For the engineer faced with the design of a superconducting magnet or some such device the engineering critical current density of the wire, $J_e$, is of paramount importance. Figure 5 shows the 4.2 K $J_e$s of the present wires plotted vs. field. The best $J_e$ of the present series is achieved by sample B3 with $1.67 \times 10^4$ A/cm$^2$ at 10 T and $0.68 \times 10^4$ A/cm$^2$ at 12 T, nearly 3.5 times higher than the highest values previously reported for IMD wires [14]. In seeking a high $J_e$, three quantities need to be maximized.

(1) First is the layer $J_c$ itself. In this regard IMD has the advantage over conventional PIT. During the HT of the PIT wire the Mg particles react into the surrounding B leaving behind pores and creating a partially connected array of MgB$_2$ macroparticles [20]. On the other hand the only porosity present in the reacted IMD wire is the axial hole – the site of the prior Mg wire – while the MgB$_2$ is present as a dense well–connected layer. Thus critical current density normalized to the area inside the chemical barrier is likely to be greater for IMD than for conventional PIT – about an order of magnitude greater – which augurs well for the $J_e$s. The choice of B powder and the possibility of including dopants in it are also important considerations.

(2) Second is the fill factor of the elemental ingredients, Mg and B. A large fill factor combined with a large $J_c$ leads directly to a high $J_e$. The present wires have powder fill factors of 19 ~ 20%, Table II.

(3) Thirdly the wires should be fully reacted. Thus strand architecture is an important consideration. According to the mechanism described in Section 3.3 it is difficult for the Mg to fully transform a thick B powder layer to MgB$_2$. Thus the 0.83 mm OD A-series wires are lower in $J_e$s than the 0.55 mm OD B-series wires although they all have the same powder fill factors. In the larger diameter wires some of the B remains unreacted even after the longest HT. The same issue has also been reported by other researchers [23]. Of course, a high $J_e$ is favored by high layer-$J_c$ combined with a fully reacted superconducting core. However, because of the reaction mechanism noted in Section 3.3, it is easier to achieve full reaction in a small diameter wire than in a thick one. On the other hand, since the critical issue is the reaction layer thickness, a better way to achieve this is ultimately a multifilament design. In either case, the shorter reaction times needed also prevent grain coarsening and its associated $J_c$ suppression.

## 4. Discussion

In order to design an MgB$_2$ strand with a high engineering $J_e$ it is necessary to combine a high layer $J_c$ with a high layer volume-fraction. In doing so the following items must be considered: (1) the choice of processing route, *ex-situ* PIT, *in-situ* PIT, or IMD; (2) the choice of B powder and the type and concentration of possible dopants; (3) the powder fill factor; (4) strand architecture; (5) HT conditions. In this context, the properties of our strands and their relationships to some of the best literature properties are summarized in terms of the 10 T critical current densities in figure 7. Table III and figure 6 also make some useful comparisons.

The optimization of PIT MgB$_2$ wire has been discussed in terms of generalized connectivity which embodies the effects of porosity and the efficiency of intergranular contact [26]. Consisting of pre-reacted MgB$_2$ powder the superconducting core of the *ex-situ* PIT wire must be less than 70% dense. The pre-reacted *in-situ* wire core also has the density of the starting powders, but after reaction further porosity develops as the Mg particles react into the surrounding B powder. Application of a "cold high pressure densification" (CHPD) step [24] densifies the starting Mg + B powder mixture but does nothing to prevent the development of porosity at the Mg-powder sites during HT. The IMD process eliminates "Mg-site porosity" as such, all of it ending up as the space previously occupied by the axial Mg rod. Since the molar volume of amorphous B is 4.59 cm$^3$/mol (hence 2B → 9.18 cm$^3$/mol) [22] and that of MgB$_2$ is 17.46 cm$^3$/mol, the expansion associated with the B itself (~ 90%), more than makes up for the preexisting porosity in the B layer (~ 40 %) and enables the resulting cylinder of MgB$_2$ to become almost fully dense and fully connected, both properties contributing to a high layer-$J_c$. The reaction Mg + 2B → MgB$_2$ takes place in three steps each of which is accompanied by expansion within the reacting B layer according to [27]: B → MgB$_7$ (18 %) → MgB$_4$ (24 %) → MgB$_2$ (28 %). Thus the overall expansion associated with the conversion of B is 89 % in agreement with previous "short-cut" (Mg + 2B → MgB$_2$) estimate.

Also contributing to the layer $J_c$ are the intrinsic properties of MgB$_2$ as influenced by the type of B powder, the presence of flux pinning additives, and the use of carbon doping. Carbon, the only element presently known to substitute into the B sublattice, increases scattering, lowers $T_c$, increases $B_{irr}$ and $B_{c2}$, reduces the anisotropy factor $\gamma = B_{c2}^{//}/B_{c2}^{\perp}$ [15], and increases high field $J_c$. Carbon-bearing compounds have been mixed in with the starting Mg + B powders to achieve *ex-situ* C doping, the most popular ones being malic acid and SiC. The non-organic additives, however, leave secondary decomposition products that could coat grain boundaries and impair connectivity and hence $J_c$. Our best results have been obtained with "SMI-boron" (from Specialty Metals Inc, U.S.A.) *in-situ* doped with C from the reduction of CH$_4$ [15-18]. In C-doped PIT strands Yang *et al*. (1, 2, 4 mol% C) [19], obtained the highest $J_c$(4.2 K, 10 T) with 4 mol% and Susner *et al.* (0-4 mol% C) [15] with 3 mol%.

Strand architecture refers to the adjustment of strand design details in order to optimize $J_e$ [19]. Thus in our present A-series and B-series strands we sought to maximize the B powder fill factor. Next, the purpose of reducing the strand diameter from 0.83 mm to 0.55 mm was to reduce the Mg-diffusion distance and hence to ensure that this enhanced fraction of B could be fully reacted in a reasonably short time, in this case 4 hours.

As stated in the introduction to this section, the essential prescription for a high $J_e$ is a combination of high layer $J_c$ with high area fraction of reacted layer. Confirmation for this is seen in Table III and figure 7. The $J_e$ is of course the product of layer $J_c$ and $MgB_2$ fill factor, curves representing which are plotted for $J_e$ values of 2 ~ 20 kA/cm$^2$. The figure then emphasizes that (i) although PIT strands may have reasonably high fill factors their inherently poor connectivity, hence low layer $J_c$, forces them onto the lower lying $J_e$ curves; (ii) although the published IMD wires [12, 14, 25] have higher connectivities and hence improved $J_c$s, they have low fill factors and correspondingly low $J_e$s (2-7 kA/cm$^2$); (iii) in the present $MgB_2$ wires, the use of IMD, associated with C doped SMI powders (hence high $J_c$), and improved strand geometry (high fill factor) produces the observed high $J_e$s (10-17 kA/cm$^2$).Thus the highest $J_e$s reported here are exhibited by the present B-series wires, and among these the best performer is sample B3 with a 4.2 K, 10 T engineering $J_e$ of 16.7 kA/cm$^2$.

## 5. Summary

By combining an IMD formation route with fine B powders, C doping, and better strand geometries, substantial increases in $J_e$ have been achieved in $MgB_2$, leading to a conductor which can be described as a second generation (2G) $MgB_2$ conductor. Monofilamentary strand designs were described, as well as the macroscopic reaction as a function of HT time. The smaller scale morphologies of various layers were also described. Finally the transport properties both in terms of layer $J_c$ and engineering $J_e$ were shown and compared to previous literature results. $J_c$ values of 1.07 x $10^5$ A/cm$^2$, and $J_e$ values of 1.67 x $10^4$ A/cm$^2$ have been achieved at 10 T and 4.2 K, and $J_e$ values of 1.0x$10^4$ A/cm$^2$ have been achieved at 11.2 T. Further optimization in $J_c$ and $J_e$ seem possible for these conductors, and multifilamentary conductors incorporating these design principles are under development.

**Acknowledgements**

This work is funded by the Department of Energy, High Energy Physics division under Grant No. DE-FG02-95ER40900, a DOE SBIR and a program from the Ohio Department of Development.

# LIST OF TABLES

**Table I: Monofilamentary Strand Diameters and HT Conditions**

| Sample name | Trace No. | Wire diameter, mm | HT time at 675 °C, h |
|---|---|---|---|
| A1 | 2712-30min-S1172 | 0.83 | 0.5 |
| A2 | 2712-1H-S1172 | 0.83 | 1 |
| A3 | 2712-8H-S1172 | 0.83 | 8 |
| B1 | 2712-30min-S1207 | 0.55 | 0.5 |
| B2 | 2712-1H-S1207 | 0.55 | 1 |
| B3 | 2712-4H-S1207 | 0.55 | 4 |

**Table II: Superconducting Layer Thicknesses, Areas and Corresponding Fill Factors**

| Sample name | MgB$_2$ layer thickness, μm | MgB$_2$ area, μm$^2$ | MgB$_2$ fill factor,%* | MgB$_2$ + B fill factor,%** |
|---|---|---|---|---|
| A1 | 0 ~ 30 | 26900 | 4.4 | 21.1 |
| A2 | 10 ~ 50 | 38200 | 6.3 | 19.7 |
| A3 | 15 ~ 60 | 44800 | 7.5 | 19.1 |
| B1 | 0 ~ 40 | 25100 | 10.1 | 19.6 |
| B2 | 0 ~ 49 | 28000 | 11.2 | 19.7 |
| B3 | 0 ~ 90*** | 46700 | 18.8 | 18.8 |

\* Area fraction of superconducting MgB$_2$ in the overall transverse cross section of the reacted wire
\** Area fraction of MgB$_2$ plus unreacted B powder after HT
\*** The B in sample B3 is fully transformed into MgB$_2$. Since the initial Mg rod is off-centered, the thickest value of MgB$_2$ layer in sample B3 reaches 90 μm

**Table III: Comparison of 10 T Engineering Critical Current Densities**

| Source | Process | Boron* | Dopants | Layer $J_c$, $10^4$A/cm$^2$ | Powder fill factor, % | MgB$_2$ fill factor, % | Engineering $J_e$, $10^3$A/cm$^2$ |
|---|---|---|---|---|---|---|---|
| *Present B-series* | | | | | | | |
| B1 | IMD | SMI | 2 mol % C | 10.7 | 19.6** | 10.1 | 9.7 |
| B2 | IMD | SMI | 2 mol % C | 10.2 | 19.7** | 11.2 | 10.2 |
| B3 | IMD | SMI | 2 mol % C | 9.3 | 18.8** | 18.8 | 16.7 |
| *Literature Results* | | | | | | | |
| [12] Li *et al.* | IMD | SMI | 2 mol % C | 10.0 | 5.2** | 5.2 | 4.6 |
| [25] Togano *et al.* | IMD | Sigma-Aldrich | 10 mol % nano-SiC | 8.7 | 4.6** | 3.2 | 2.9 |
| [14] Ye *et al.* | IMD | Sigma-Aldrich | 10 mol % nano-SiC + aromatic hydrocarbon | 4.8 | 14.0** | 7.7 | 3.7 |
| [12] Li *et al.* | PIT | SMI | 2 mol % C | 1.4 | 18.1*** | 18.1 | 2.6 |
| [24] Hossain *et al.* | PIT + CHPD | SB 99 | 10 wt % C$_4$H$_6$O$_5$ (malic acid) | 2.7 | 23.7*** | 23.7 | 6.4 |

\* SMI: plasma synthesized B from Specialty Materials Inc. (SMI) [15-19]; Sigma-Aldrich: high purity amorphous B powder (99.99%, 300 mesh, Sigma-Aldrich Co.); SB 99: high purity amorphous B powder (99%, less than 50 ppm metallic contamination, average particle size of 60 nm, SB Boron Corp.)
\*\* Area fraction of MgB$_2$ plus unreacted B after HT (IMD)
\*\*\* Area fraction of reacted powder in the core of the PIT wire

**LIST OF FIGURES**

**Figure 1:** Back scattered SEM images of (a) sample A1 (0.83 mm OD, HT 675 °C/30 min), (b) sample A2 (0.83 mm OD, HT 675° C/1 hour), and (c) sample A3 (0.83 mm OD, HT 675 °C/8 hours).

**Figure 2:** High resolution fracture-SEM images of sample A2 at (a) B rich layer, and (b) $MgB_2$ layer.

**Figure 3:** Back scattered SEM images of (a) sample B1 (0.55 mm OD, HT 675 °C/30 min), (b) sample B2 (0.55 mm OD, HT 675 °C/1 hour) and (c) sample B3 (0.55 mm OD, HT 675 °C/4 hours).

**Figure 4:** Layer $J_c$ vs. $B$ curves of A-series and B-series wires at 4.2 K.

**Figure 5:** Engineering $J_e$ vs. $B$ curves of A-series and B-series wires at 4.2 K.

**Figure 6:** Comparison of the engineering $J_e$ vs. $B$ curves of the B-series wires with literature results. In presenting this comparison we note that the wire of reference [25] is multifilamentary and as such may have a smaller filling factor than the other monocore strands and a correspondingly reduced $J_e$.

**Figure 7:** Relationship between engineering $J_e$, layer $J_c$ and $MgB_2$ fill factor of different wires at 4.2 K, 10 T.

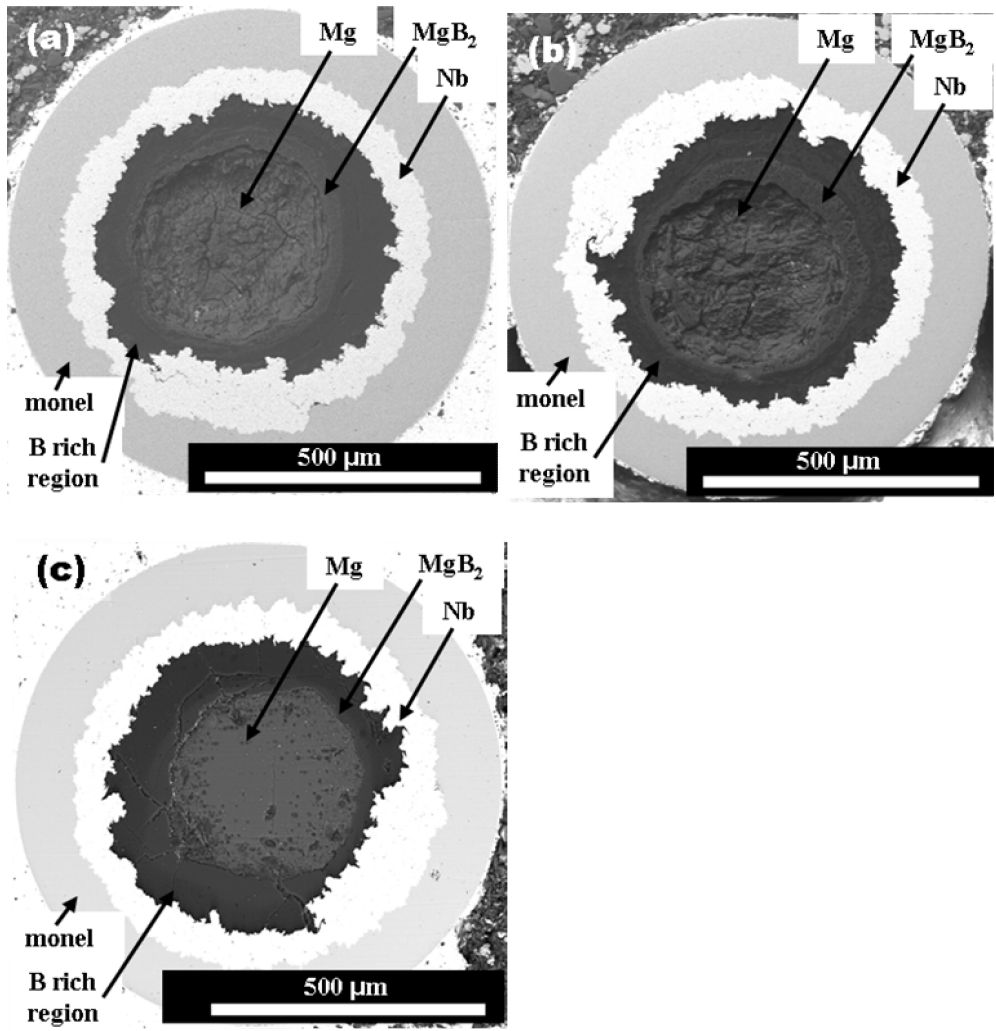

**Figure 1:** Back scattered SEM images of (a) sample A1 (0.83 mm OD, HT 675 °C/30 min), (b) sample A2 (0.83 mm OD, HT 675° C/1 hour), and (c) sample A3 (0.83 mm OD, HT 675 °C/8 hours).

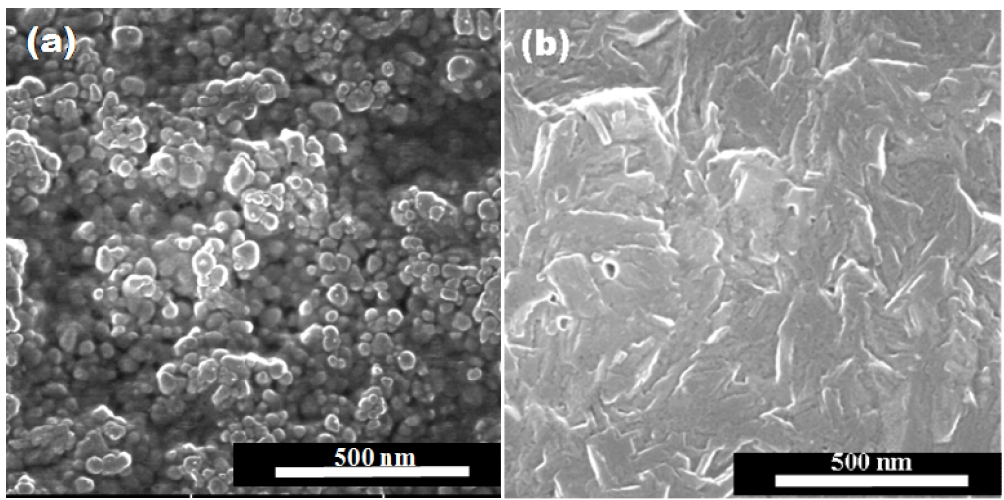

**Figure 2:** High resolution fracture-SEM images of sample A2 at (a) B rich layer, and (b) MgB$_2$ layer.

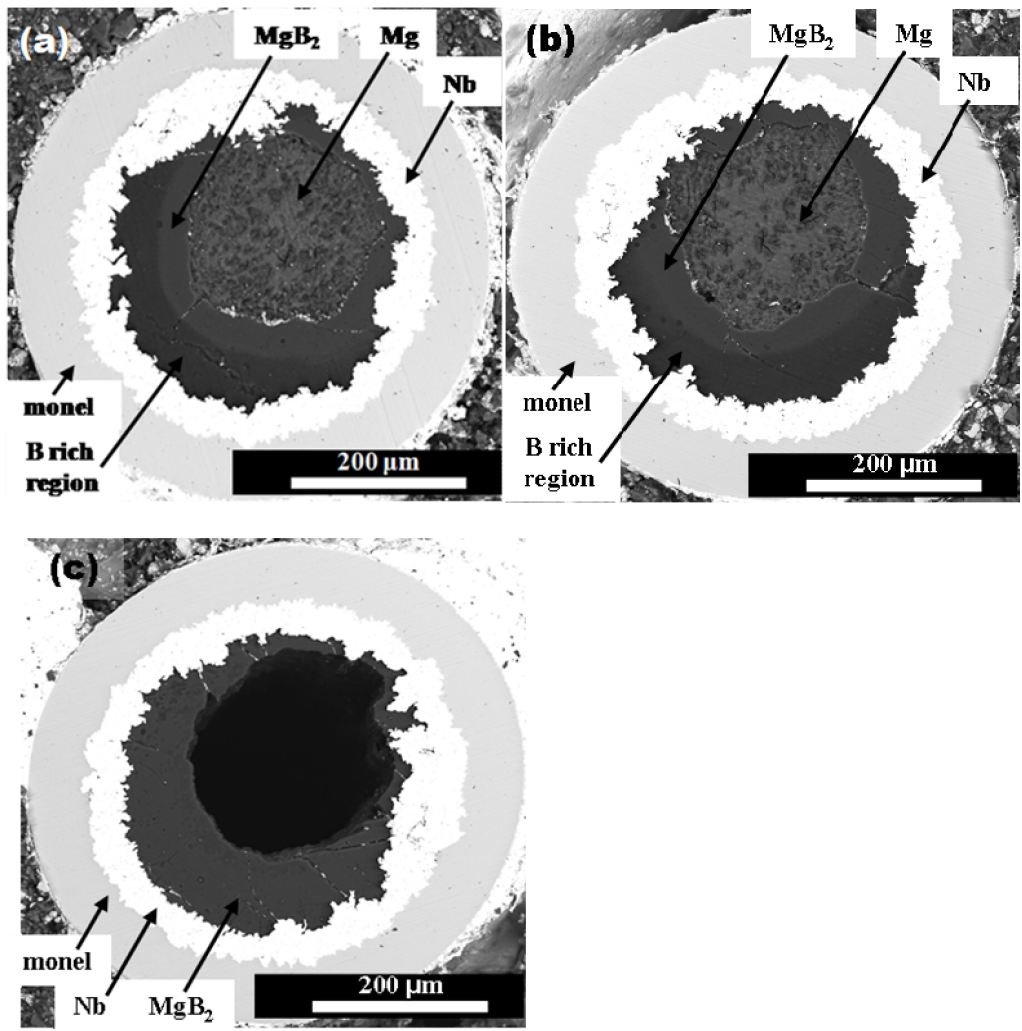

**Figure 3:** Back scattered SEM images of (a) sample B1 (0.55 mm OD, HT 675 °C/30 min), (b) sample B2 (0.55 mm OD, HT 675 °C/1 hour) and (c) sample B3 (0.55 mm OD, HT 675 °C/4 hours).

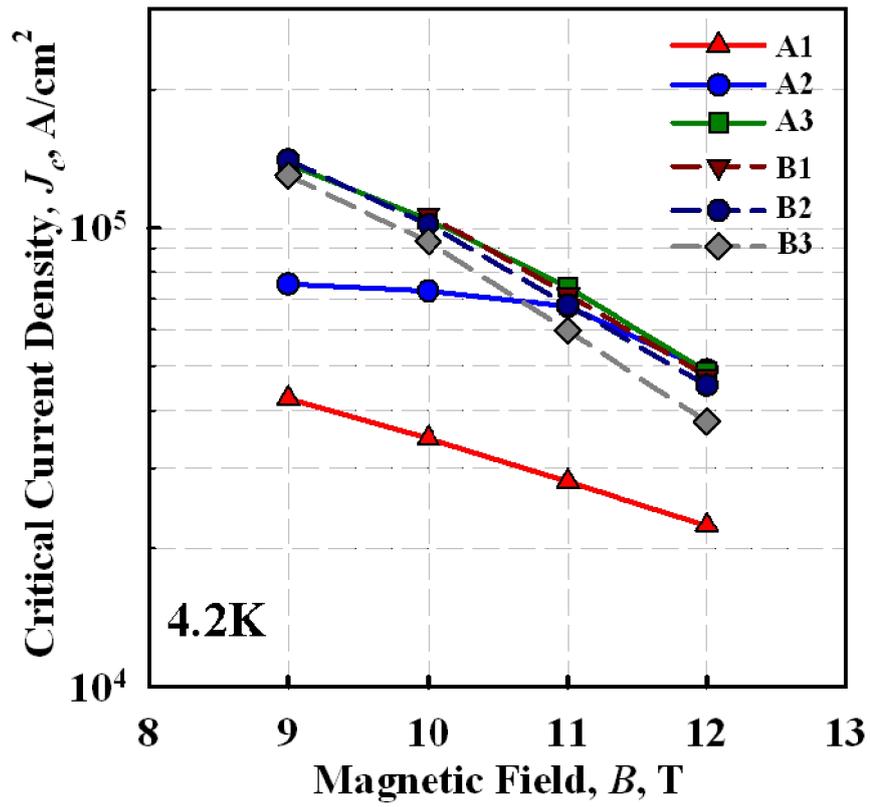

**Figure 4:** Layer $J_c$ vs. $B$ curves of A-series and B-series wires at 4.2 K.

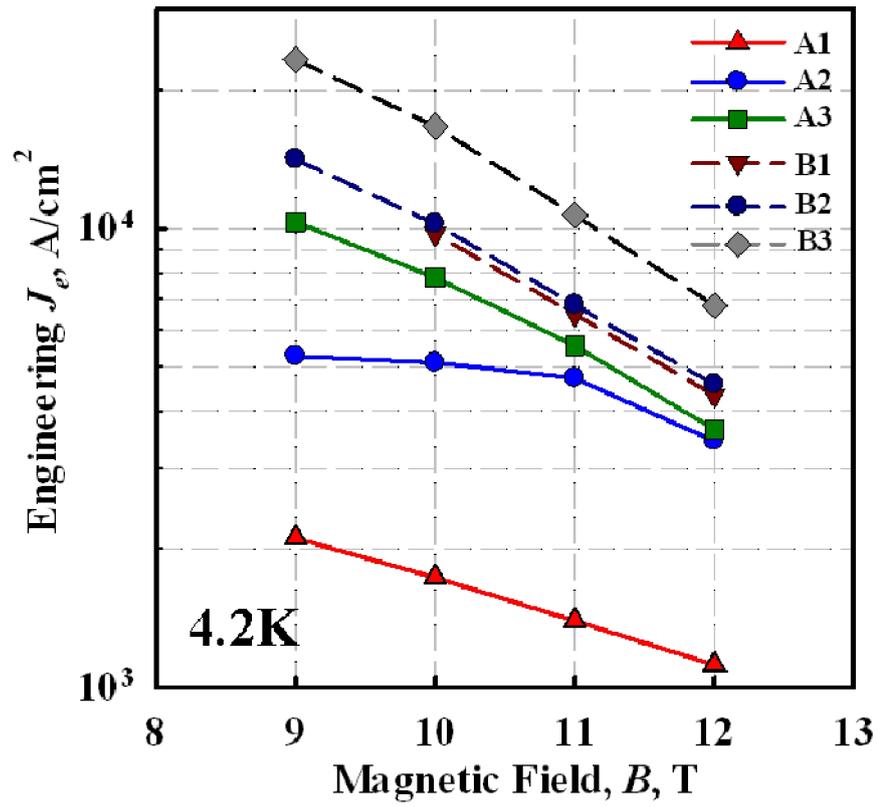

**Figure 5:** Engineering $J_e$ vs. $B$ curves of A-series and B-series wires at 4.2 K.

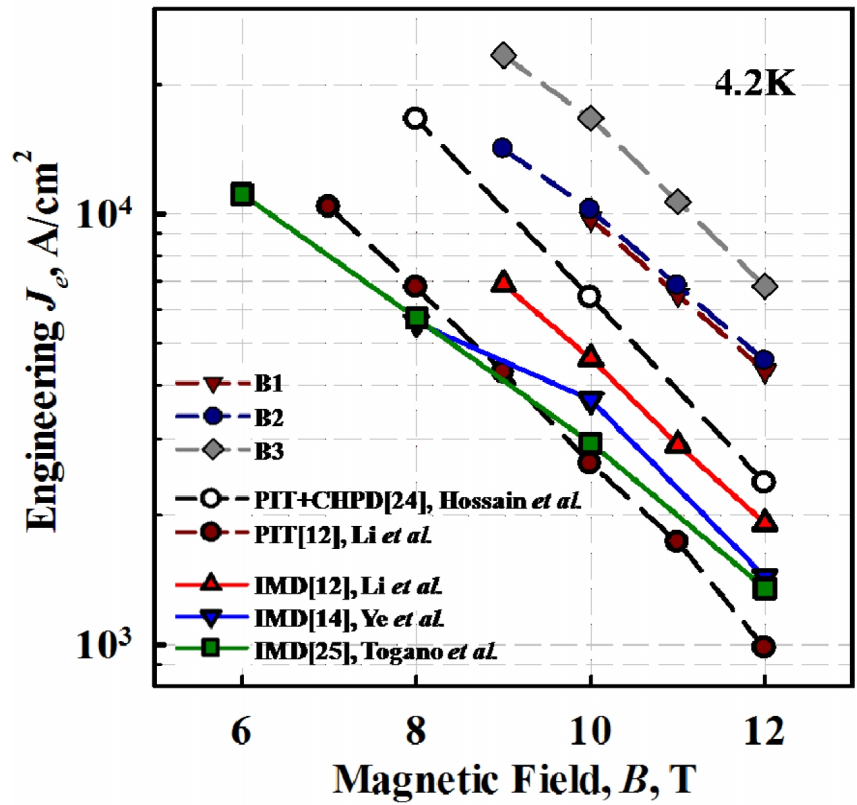

**Figure 6:** Comparison of the engineering $J_e$ vs. $B$ curves of the B-series wires with literature results. In presenting this comparison we note that the wire of reference [25] is multi-filamentary and as such may have a smaller fill factor than the other monocore strands and a correspondingly reduced $J_e$.

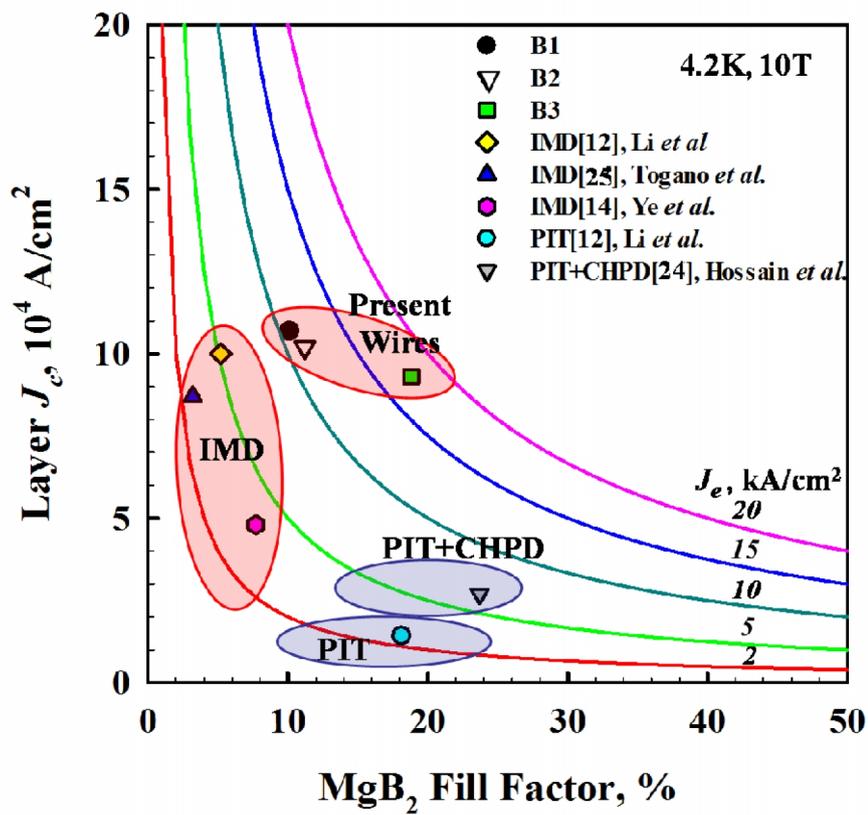

**Figure 7:** Relationship between $J_e$, layer-$J_c$ and $MgB_2$ fill factor of different wires at 4.2 K, 10 T.